# Infection-responsivity of Commercial Dressings Through Halochromic Drop-casting

Charles Brooker[1,2 a)] and Giuseppe Tronci[1,2 b)]

[1] *Clothworkers' Centre for Textile Materials Innovation for Healthcare (CCTMIH), School of Design, University of Leeds, Leeds LS2 9JT, UK*
[2] *School of Dentistry, St. James's University Hospital, University of Leeds, Leeds LS9 7TF, UK*

[a)] *Corresponding author: c.brooker@leeds.ac.uk*
[b)] *g.tronci@leeds.ac.uk*

**Abstract.** Infection control remains one of the most challenging tasks in wound care, due to growing antimicrobial resistance and ineffective infection diagnostic tools at the point-of-care. To integrate therapeutic wound dressings with wound monitoring capability at the point-of-care to enable informed clinical decision-making, we investigate the encapsulation of a halochromic dye, i.e. bromothymol blue (BTB), onto two commercial dressings, i.e. Aquacel® Extra™ and Promogran®, through a simple drop-casting method. Our concept leverages the infection-associated rise in wound pH, on the one hand, and BTB's colour change capability in the pH range of healing (pH: 5-6) and infected wounds (pH > 7), on the other hand. BTB-encapsulated samples show a prompt colour switch (yellow/orange → blue) following 1-hour incubation at pH 8. The effect of swelling ratio, chemical composition and microstructure is then explored to draw relationships between colour change capability and dressing dye retention.

## INTRODUCTION

Wound dressings play a crucial role in the management of non-self-healing, i.e. chronic, wounds, aiming to create an optimal moist environment for tissue repair. They protect wounds from contaminants, manage exudate, and provide physical support. The development of wound dressings has been influenced by advancements in materials science, biomaterials, and our understanding of wound healing [1].

One of the main barriers to chronic wound healing is the risk of recurrent infection, which is exacerbated by the growing trends in antimicrobial resistance [2] and the lack of infection diagnostic tools. Consequently, the delayed healing of chronic wounds in the United Kingdom costs the National Health Service (NHS) £5 billion annually, driven by expenditure associated with extended hospital stays, prolonged treatment periods, and the occurrence of clinical complications, including gangrene and the need for amputation [3].

The therapeutic efficacy of hydrogel-based dressings for chronic wounds depends on their absorption capacity and their ability to maintain an optimal moisture balance [4]. Inadequate absorption of wound exudate can lead to fluid accumulation and the risk of maceration, while strong adherence to the wound bed can lead to pain during dressing changes [5]. In addition to fostering a pro-healing milieu in the chronic wound, additional functionalities should be integrated into current therapeutic dressings, aiming to easily detect infection, and ensure prompt variations in clinical and home care. Despite this urgent need, infection diagnosis still predominantly relies on clinical assessment, fostering risks of suboptimal therapeutic regimens, antibiotic misuse, and antibiotic-resistant infections [6].

Current commercial dressings have functional limitations in real-time monitoring of wound parameters, which can enable proactive management of infection and prompt therapies. There are several biomarkers which could be used for wound monitoring and infection detection, with wound pH a promising infection biomarker given the alkaline shift that takes place following infection onset [7,8].

Exploiting the colour change capability of halochromic dyes, this work investigates the point-of-care drop casting integration of two commercial dressings, i.e. Aquacel® Extra™ and Promogran®, with bromothymol blue. The drop-casting method offers clinicians a straightforward way to integrate the dye into the dressing at the point-of-care, resulting in a multi-functional dressing when necessary, while simultaneously reducing fabrication costs and regulatory burden. Our objectives aim to accomplish a visual signal at an infection-associated wound environment (pH > 7), on the one hand, and to quantify the dye retention in the dressing, on the other hand, to assess response durability and minimise the impact of dye release on the wound environment.

## MATERIALS AND METHODS

BTB was purchased from Alfa Aesar (Heysham, UK), dressings of Aquacel® Extra™ and Promogran® were purchased from EasyMeds Pharmacy (Huddersfield, UK), and phosphate-buffered saline (PBS) was purchased from Lonza (Slough, UK). All other chemicals were purchased from Sigma-Aldrich unless specified.

### Preparation of Drop-cast Dressings

After thorough vortexing, a drop of up to 100 µl of BTB solution (0.2 wt.% BTB in deionised water) was applied to the dry dressings and exposed to air for 30 min.

### Swelling Tests

Dry samples of known mass ($m_d$, n=7) were individually incubated in either distilled water or PBS (10 mM, pH 7.4) at 25 °C for 24 hours. The swelling ratio ($SR$) was calculated after 24 hours according to Equation 1, as previously reported [9]:

$$SR = \frac{m_s - m_d}{m_d} \times 100 \tag{1}$$

where $m_s$ is the swollen mass of collagen hydrogel samples.

### Quantification of BTB Loading

To quantify the BTB loading content of drop-cast samples, a gravimetric method was employed (n=7). The mass of the individual samples ($m_i$) was recorded using a precision balance before the BTB solution (100 µl, 0.2 wt.% BTB) was drop-cast using a micropipette. After air-drying the samples for 48 hours, the final mass ($m_f$) was recorded using a precision balance and the loading efficiency ($LE$) was calculated using Equation 2:

$$LE = \frac{(m_f - m_i)}{m_{BTB}} \times 100 \tag{2}$$

where $m_{BTB}$ is the mass of BTB contained in the volume of the aqueous solution applied to the samples.

### Dye Release Measurements

Dye retention was indirectly assessed by measuring the release of the dye out of the dressings. Individual samples (n=3) containing up to 200 µg of BTB were incubated at room temperature in 5 mL of McIlvaine solution adjusted to either pH 5 or pH 8. Samples were placed in Petri dishes (Corning, Corning, NY, USA) onto a layer of McIlvaine solution, simulating a dressing placed on a moist wound. The amount of BTB released from the dressing onto each solution after 1-hour incubation was determined via UV-Vis spectrophotometry. Calibration curves were built with McIlvaine solutions adjusted at pH 5 and pH 8 loaded with varying amounts of BTB, with recordings taken at 432 nm and 616 nm, respectively. Resulting absorbance data were used to derive the amount of BTB released from the samples at each time point and solution pH.

### Scanning Electron Microscopy

Native and drop-cast samples of Aquacel® Extra™ and Promogran® were inspected via scanning electron microscopy (SEM) using a Hitachi S-3400N microscope (Hitachi, Tokyo, Japan). Before examination, all samples were gold-sputtered using an Agar Auto sputter coater (Agar Scientific, Stansted, UK). The SEM was fitted with a tungsten electron source and the secondary electron detector was used. The instrument was operated with an accelerating voltage of 3 kV in a high vacuum with a nominal working distance of 10 mm. Fibre and pore diameters were recorded manually using ImageJ software, with 80 repeats.

## Statistical Analysis

Data normality tests were carried out using OriginPro 8.51 software (OriginPro, OriginLab Corporation, Northampton, MA, USA). Statistical differences were determined by one-way ANOVA and the post hoc Tukey test. A *p*-value lower than 0.05 was considered significantly different. Data are presented as mean ± SD.

## DISCUSSION

Two commercial wound dressings were selected to investigate the infection responsivity following drop-casting of the halochromic dye. Aquacel® Extra™ is a commercially available dressing which is comprised of hydrogel-forming sodium carboxymethylcellulose (Na-CMC) fibres; as such, it is often used to manage moderate to highly exuding wounds. Promogran® is another commercially available dressing which is comprised of 55% collagen and 45% oxidised regenerated cellulose (ORC) in a spongy, freeze-dried matrix, and is mainly applied for the management of chronic and acute wounds. The electron microscopy images of these two dry dressings are presented in Figure 1, where two distinct microstructures can be observed, i.e. the Na-CMC fibres of Aquacel® Extra™ (Figure 1a) and the porous matrix of Promogran® (Figure 1b). The Na-CMC fibres of Aquacel® Extra™ had uniform diameters (Ø = 11 ± 1 µm) and the average pore size of the Promogran® matrix was calculated to be 145 ± 42 µm (Table 1).

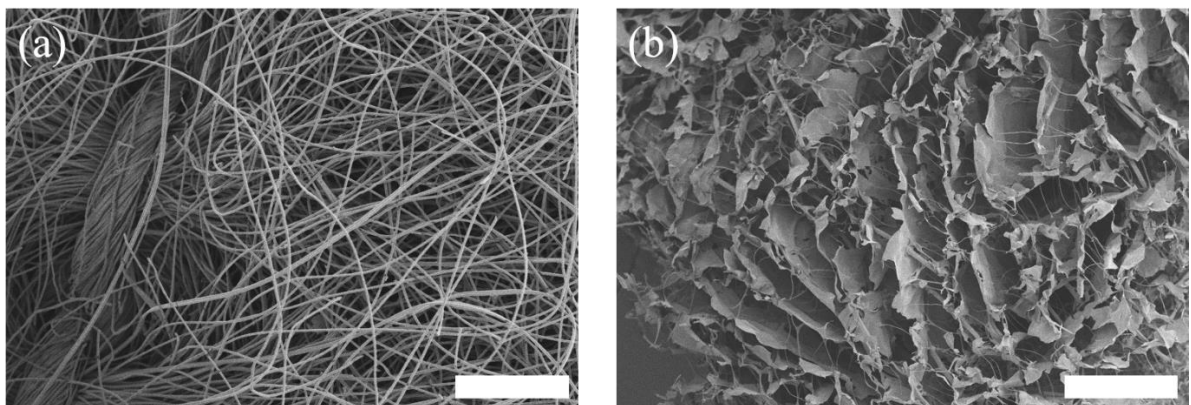

**FIGURE 1.** SEM micrographs of (a) native Aquacel® Extra™; (b) native Promogran®. Scale bars = 500 µm.

To quantify their wound exudate management capability, the liquid adsorption of the two dressings was recorded after a 24-hour incubation in both distilled (DI) water and PBS (10 mM, pH 7.4). Unsurprisingly, both dressings showed a remarkable swelling. Promogran® had a higher averaged swelling ratio compared to Aquacel® Extra™ in both DI (*SR*: 1711–2194 wt.%, Table 1) and PBS (*SR*: 1686–2100 wt.%, Table 1). Though no statistical significance was recorded, the increased swelling measured on Promogran® compared to Aquacel® Extra™ could be attributed to the presence of the two hydrophilic biopolymers, i.e. collagen and ORC, and porous structure in the former dressing.

**TABLE 1.** Dry-state diameters (Ø, n=80) of fibres (Aquacel® Extra™) and pores (Promogran®), and swelling ratio (*SR*) measured after 24 hour-incubation in DI water and PBS (n=7) of commercial dressings.

| Dressing | Ø (µm) | SR (wt.%) | |
|---|---|---|---|
| | | DI | PBS |
| Aquacel® Extra™ | 11 ± 1 | 1711 ± 130 | 1686 ± 140 |
| Promogran® | 145 ± 42 | 2194 ± 181 | 2100 ± 144 |

As mentioned in the introduction, an additional dressing functionality was introduced by exploiting the colour change capability of BTB, following on from earlier successful investigations via drop casting and electrospinning [10,11]. After the addition of the halochromic dye via a drop-casting method, the loading efficiency ($LE$) was measured to confirm encapsulation of BTB in the dressings. High insignificantly different values were recorded on Aquacel® Extra™ ($LE$= 99 ± 2 wt.%) and Promogran® ($LE$= 99 ± 3 wt.%), supporting the validity of the drop-casting method. The microstructure of the drop-cast dressings was therefore analysed using SEM (Figure 2) to assess any effect of the dye encapsulation. In the case of Aquacel® Extra™, in the region where the dye was added through liquid injection, the solution has been absorbed directly into the sodium carboxymethylcellulose fibres, which coalesced to form a cohesive gel. Similarly, where the dye was added to Promogran®, the porous structure of the dressing collapsed, indicating that the water-induced swelling irreversibly affects the microstructure of the dressing. The introduction of covalent crosslinks at the molecular scale could be pursued aiming to retain the original configuration, as previously observed with electrospun gelatin fibres [12] and hyaluronic acid-based hydrogels [13].

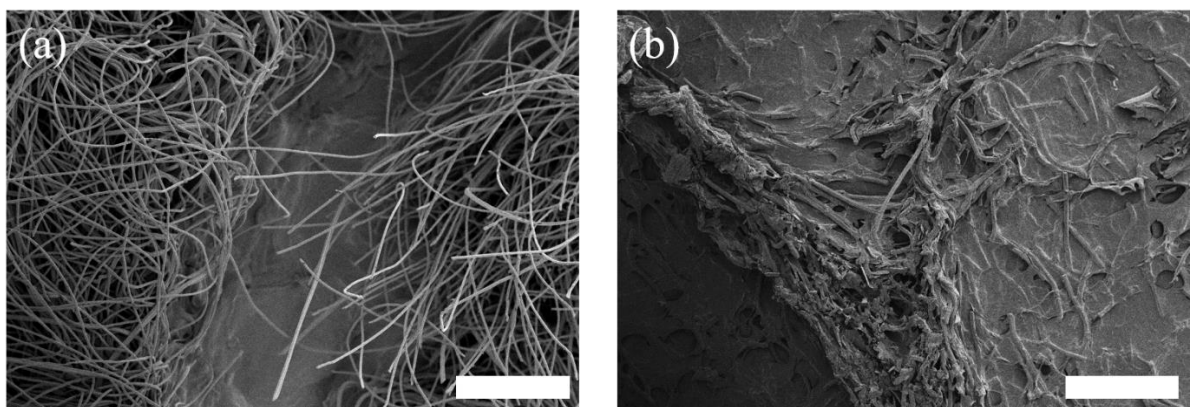

**FIGURE 2.** SEM micrographs of BTB drop-cast dressings: (a) Aquacel® Extra™; (b) Promogran®. Scale bars = 500 µm.

After thirty minutes of air-drying, the drop-cast dressings were added to a layer of pH 5 and 8 McIlvaine solution, to mimic the pH of a healthy and infected wound, respectively. The dye release away from, and the colour change of, the dressing was quantified using digital macrographs and UV-vis spectroscopy, respectively. These two investigations were deemed relevant aiming to examine the durability of the infection response of the commercial dressings. The dye release after 1 hour is stated in Table 2. At pH 5, the dye release from Promogran® is lower than the release from Aquacel® Extra™ (2 ± 0 vs 7 ± 1 wt.%), however, at pH 8, the dye release from Promogran® is higher than the release from Aquacel® Extra™ (12 ± 1 vs 5 ± 1 wt.%). The cytotoxicity of the dye released from a drop-cast collagen dressing was previously assessed against L929 fibroblast cells, revealing an average cell viability of over 90% after 7 days [10]. The aforementioned extract contained a BTB content more than three times higher than that released from the commercial dressings selected in this study, supporting the cell tolerability of BTB in this release range and indicating no detrimental impact on the wound environment.

**TABLE 2.** Dye release from commercial dressings following 1-hour incubation in pH 5 and 8 McIlvaine solutions.

| Dressing | Dye Release (wt.%) | |
|---|---|---|
| | pH 5 | pH 8 |
| Aquacel® Extra™ | 7 ± 1 | 5 ± 1 |
| Promogran® | 2 ± 0 | 12 ± 1 |

The difference in release rates between the dressings at different pH values can be explained by the chemical composition of the dressings and the chemical configuration of the dye itself. Promogran® is comprised of 55% collagen and 45% ORC, both of which contain ionisable residues in the form of carboxylic and primary amino groups, on the one hand, and carboxylic acid groups only, on the other hand. Likewise, BTB is a pH-sensitive compound that presents one negative charge in acidic environments and two negative charges in alkaline environments (pH> 7). At pH 5, the drop-cast sample of Promogran® contains BTB with a monovalent charge as well as free positively charged lysines of collagen and negatively charged carboxylic groups associated with the ORC. Therefore, there is an increased electrostatic interaction between dye and collagen but a decreased electrostatic repulsion between dye and ORC.

At pH 8, the chemical structure of the drop cast sample is altered in that BTB has a bivalent charge and there is also a decreased amount of protonated collagen lysines ($pK_a$ ~10.5), while ORC's deprotonated carboxyl groups remain unaffected ($pK_a$ ~ 4). Therefore, there is a decreased electrostatic interaction between dye and collagen but an increased electrostatic repulsion between dye and ORC, which explains the increased release of dye observed when the dressing is incubated in the alkaline environment.

Aquacel® Extra™ is comprised of Na-CMC fibres which contain carboxylic groups. At pH 8, we see a more rapid dye release rate. This observation could be explained by the increase in negative charges on the dye, and the deprotonation of the carboxylic groups, suggesting an increased electrostatic repulsion between the dye and the carboxyl groups. At pH 5 the release rate is slower as the charge on the dye is -1, therefore there is a decrease in electrostatic repulsion.

When comparing the release rates of Aquacel® Extra™ with Promogran®, at pH 8, the dye released from Promogran® following 1-hour incubation is more than double that recorded from Aquacel® Extra™. This could be explained by the increased swelling ratio measured on the former samples (Table 1), which enables increased diffusion of the dye from the dressing. At pH 5, on the other hand, there is a decreased dye release from Promogran® compared to Aquacel® Extra™. This is in line with the effect of protonated lysines enabling electrostatic interactions between the dye and collagen.

Figure 3 displays a matrix of photographs of the drop-cast dressings, contrasting the dye colour, and spread of dye through Aquacel® Extra™ and Promogran® over one hour at both pH 5 and pH 8. The colour change from yellow/orange to blue occurs almost instantaneously when the dressing is added to the pH 8 medium. This colour shift is associated with the change in the molecular configuration of the dye. Below pH 7, BTB presents a monovalent anion with the sulfonate group; however, above pH 7, proton dissociation from the phenolic group results in a bivalent anion and an increased negative electrostatic charge. The photographs in Figure 3 also show varying degrees of spreading of the dye, with the Aquacel® Extra™ dressings displaying slightly more dye spread after 1 hour.

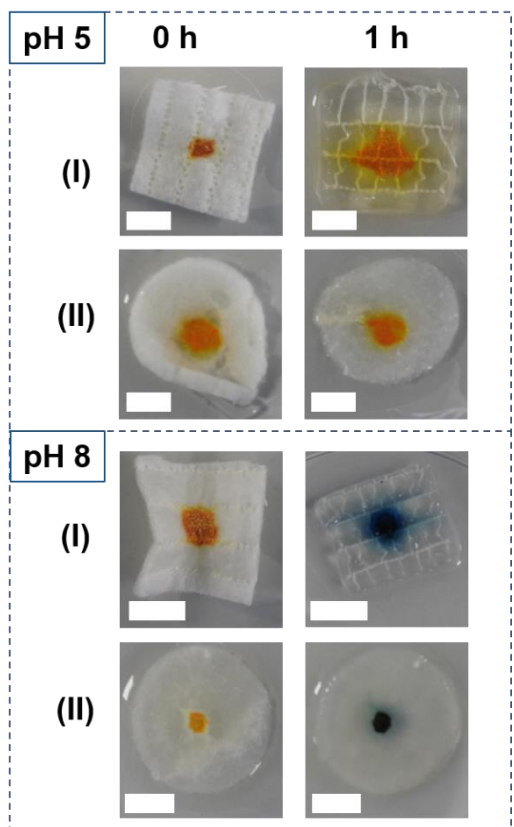

**FIGURE 3.** Matrix of photographs of the BTB drop-cast dressings immediately after drop-casting (t=0h) and following 1-hour incubation (t=1h) in pH 5 and pH 8 McIlvaine solution. (I) Aquacel® Extra™; (II) Promogran®. Scale bars = 1 cm.

# CONCLUSION

Encapsulation of BTB, a halochromic dye, was realised through a simple drop-casting method onto two commercial dressings, Aquacel® Extra™ and Promogran®. Analysis via SEM demonstrated that the dressing microstructure changed after drop-casting, with the Na-CMC fibres of Aquacel® Extra™ coalescing to form a cohesive gel, and the porous matrix of Promogran® collapsing. These dressings subsequently displayed a rapid colour change upon incubation at pH values associated with infected wounds (pH> 7). Dye release rates from Promogran® following 1-hour incubation at pH 8 were more than double that recorded from Aquacel® Extra™, which can be explained by the increased swelling ratio measured on the former samples. Whereas, at pH 5, there was a decreased dye release from Promogran® compared to Aquacel® Extra™, which is attributed to electrostatic interactions between the dye and protonated lysines of collagen. Overall, these investigations present a simple and rapid strategy to integrate commercial therapeutic dressings with wound monitoring capability at the point-of-care, with no impact on dressing manufacture and regulatory compliance. While the dye loading efficiency was 99 wt.% in both dressings and the dye remains largely confined to the structure following 1-hour incubation (dye release $\leq$ 12 wt.%), the dye retention capability of dressings should be investigated at increased time points (e.g. days) to ensure long-lasting responsivity and compliance with wound care applications.

# ACKNOWLEDGEMENTS

The Clothworkers' Company (London, UK) is gratefully acknowledged for financial support. The authors gratefully acknowledge Michael Brookes for technical assistance with SEM.